\documentclass[nofootinbib]{revtex4-1} 
\usepackage[utf8]{inputenc} 
\usepackage{amsmath} 
\usepackage{graphicx} 
\usepackage{subcaption} 
\usepackage{amssymb} 
\usepackage{color} 
\usepackage{cancel} 
\usepackage{framed} 
\usepackage{comment} 
\usepackage[margin=0.7in]{geometry} 
\usepackage{mathtools} 
\usepackage{hyperref} 
\usepackage{chngcntr} 
\usepackage{soul}
\usepackage[toc,title,page]{appendix} 

\counterwithin{equation}{section}

\begin{document}

\title{WKB approximation to bosonic dark matter}

\author{Lauren Street} 
\email{streetlg@mail.uc.edu} 
\affiliation{Department of Physics, University of Cincinnati}

\author{Peter Suranyi} 
\email{peter.suranyi@gmail.com} 
\affiliation{Department of Physics, University of Cincinnati}

\author{L.C.R. Wijewardhana} 
\email{rohana.wijewardhana@gmail.com} 
\affiliation{Department of Physics, University of Cincinnati}

\begin{abstract} 
Galactic dark matter halos may be composed of ultralight axions (ULAs) ($m_a \lesssim 1$ eV) with wave functions that satisfy nonlinear Schr\"{o}dinger-Poisson equations (SPA). We find eigenstates of SPA in WKB approximation.  The expansion parameter of the WKB approximation is $\delta=1/\sqrt{S}$, where $S=2 M R G m_a^{2}$, with $M$ being the total mass, $R$ the radius of the halo, and $G$ the gravitational constant.
$S\gg 1$ for almost all galaxies, even if the ULA mass is as small as $m_a=10^{-22} $ eV, making the leading order WKB approximation almost exact. As the level spacing of bound states is roughly proportional to $\delta$, the number of states in the gravitational well is huge. We do not see a reason why not all or most of them contribute to the halo. Using an appropriate distribution function allows the summation of states to construct the profile of the halo as a function of the gravitational potential, which can be found solving the Poisson equation. Using various energy distribution functions, we obtain results similar to those in simulations. Future plans include investigations of collapse through time dependent generalizations, and inclusion of self-interactions, which also induce decay processes of the halo.
\end{abstract}

\maketitle

The structure of galaxies and the rotation curves of stars in galaxies can potentially be explained with the assumption that most of galactic matter is composed of presently unknown particles, termed dark matter (DM), which interact very weakly with particles of the Standard Model. One of the most popular variants of DM is the weakly interacting massive particle (WIMP), consisting of massive, non-relativistic particles, heavier than neutrinos~\cite{1982ApJ...263L...1P,Bond:1982uy,Blumenthal:1982mv}. Since no such particles, in the appropriate mass range, have been discovered yet, other alternatives for DM have also been considered. Among others, a prominent candidate is ultralight axions (ULAs) with Compton wavelengths ranging from cosmic size~\cite{Baldeschi:1983mq,Turner:1983he,Press:1989id,Sin:1992bg,Hu:2000ke,Guzman:2004wj,Boehmer:2007qa,Weinberg:2013aya,Marsh:2015wka,Hui:2016ltb,Lee:2017qve} to that of masses $m_a  \sim 1$ eV \cite{Ferreira:2020fam}.

There have been many simulations of the collapse of ULAs on galactic scales without \cite{Schive:2014dra,Schive:2014hza,Schwabe:2016rze,Veltmaat:2016rxo,Du:2016aik,Mocz:2017wlg,Levkov:2018kau} and with baryonic feedback \cite{Mocz:2019pyf}. In both cases, ULA systems were shown to collapse to a condensed core, surrounded by a virialized halo of non-relativistic ULA. Recently, there have also been simulations performed for ULAs with self-interactions, \cite{Amin:2019ums,Glennon:2020dxs,Mocz:2023adf}, for systems composed of multiple flavors without self-interactions \cite{Huang:2022ffc}, and for systems composed of multiple flavors with self-interactions \cite{PhysRevD.107.063520}. In order to study dynamic perturbations to the central soliton, multiple eigenstates  of a Schrodinger Poisson system 
where the gravitational potential for higher modes is generated by the solitonic ground state was performed
 in reference ~\cite{Zagorac:2021qxq}.
These states can be used to construct halos of low mass galaxies where due the the low value of S only a small number of states contribute to the halo.

In another work \cite{Lin:2018whl}, self-consistent simulations of the halo constructed from excited states of the Schr\"odinger-Poisson (SP) equations were performed. The authors found that the collapse of the system consisted of a condensed soliton core surrounded by a halo composed of excited eigenstates. In a subsequent work \cite{PhysRevD.101.081302}, systems satisfying SP equations were analyzed assuming composition of a small number of energy eigenstates, including the stability and virialization of the system.

The purpose of this work is to construct DM from self-adjoint or complex ULAs with self-interactions using WKB approximation. We ignore self-interactions in solving the equations of motion, but in a subsequent work we will consider the effect of $2 \rightarrow 2$ interactions on the stability of excited eigenstates. We emphasize that, because only this particular interaction is relevant, our model can be used for real, or just as easily for complex, scalar fields. For the sake of simplicity, we focus on a real scalar field giving rise to a ULA subject to a $\Phi^4$ self-interaction. Such an interaction is the leading order expansion term of an axion potential, $V=m_a^2f^2[1-\cos(\Phi/f)]$. For the range of galactic sizes and ULA masses considered here, the contribution of self-interaction terms to the equations of motion is negligible compared to that of the gravitational interaction. The ratio of self-interactions to gravitational interactions scales as 
\begin{equation*} 
\frac{\text SI}{GI}\sim \frac{M_P^2}{f_a^2}\frac{1}{m_a^2R^2},
\end{equation*} 
where $M_P = G^{-1/2}$ is the Planck mass and $G$ is Newton's constant, $f_a$ is the axion decay constant, $R$ is the radial scale of the system and $m_a$ is the mass of the axion. Using $m_a=10^{-22}eV$, $f/M_p=10^{-3}$ for the Milky Way with $R=10^5$ light year, we obtain $SI/GI\simeq 10^{-4}$.   

Self-interactions of ULAs may be important for extremely small galaxies. In fact, based on studies of axion stars~\cite{PhysRevD.84.043531,Chavanis:2011zm,Eby:2015hyx} they can possibly generate a cutoff in the mass of small, stable galaxies with very large densities. That possibility will be investigated in future publications. 

\section{Equations of motion} 
Our basic assumption is that radial eigenfunctions of the halo satisfy a Schrodinger-Gross-Pitaevskii equation 
\begin{align}
\label{scheq} E_{nl}\psi_{nl}=-\frac{1}{2m_a}\left[\psi_{nl}''+\frac{2}{r}\psi_{nl}'\right]+\left[\frac{1}{2 m_a}\frac{l(l+1)}{r^2}+V_g\right]\psi_{nl}, 
\end{align} 
where quantum numbers $n$ and $l$ characterize eigenstates, and $V_g$ is the gravitational potential. The most general stationary state wave function of the halo is 
\begin{align}\label{totalwave} \Psi({\bf r},t)=\sum_{nlm}\psi_{nl}(r) Y_{lm}(\theta,\phi)e^{i(E_{nl}t+\,\alpha_{nlm})}, 
\end{align} 
where $\alpha_{nlm}$ are random phases. Then the gravitational potential is 
\begin{align} \label{Vg}
V_g&=- G \, m_a \int d^3r'\frac{\Psi(\bf{r}')^2}{|\bf{r}-\bf{r}'|} \simeq- G \, m_a\sum_{nl}(2l+1) \int d^3r'\frac{\psi_{nl}(r')^2}{|\bf{r}-\bf{r}'|}, 
\end{align} 
where we average over time and random phases, and assume spherical symmetry.

The main result of this paper is the analytic derivation of the density distribution of a ULA halo. Suppose the gravitational potential of the system, $V_g(r)$, is known. Then the density distribution satisfies the Poisson equation 
\begin{equation}\label{poisson} 
\nabla^2V_g(r)=-4\pi\,G\,m_a \rho(r). 
\end{equation} 
If we can construct the density distribution as a function of $V_g$, then Eq. (\ref{poisson}) constitutes a second order differential equation for $V_g$ which can be solved, providing the density distribution.  This density distribution can then be compared to simulations or observations.

Expanding the wave function in radial coordinates, assuming again that interference terms are negligible when taking averages and the density distribution is spherically symmetric, the number density can be written as 
\begin{equation} 
\rho(r)=m_a \sum_{nl} (2l+1)|\psi_{nl}|^2, 
\end{equation} 
where we normalize wave functions as 
\begin{equation} 
\int d^3r|\psi_{nl}|^2 = N_{nl}, 
\end{equation} 
where $N_{nl} \equiv M_{nl}/m_a$ and $M_{nl}$ is the total mass of states having quantum numbers $n$ and $l$. 

In phenomenological models (\cite{Navarro:1995iw,Burkert:1995yz}) the central density and radial scale are undetermined free parameters. To compare various radial scales we introduce the universal radial scale parameter $R$, the rescaled dimensionless coordinate, $z=r/R$, and the gravitational scaling function, $w(z)$, through the equation 
\begin{align}\label{gravpot} 
V_g (z)=-G\frac{m_a^2}{R}\int d^3z'\frac{\tilde\rho(z')}{|z-z'|}=-\frac{GM\, m_a}{R}w(z), 
\end{align} 
where we rescale the density as $\tilde\rho(z) = \left(R^3/m_a\right) \rho(r)$. Choosing $R$ as the harmonic average of $r$ weighted over the density, 
\begin{align}\label{harmonic} 
\frac{1}{R} = \left\langle \frac{1}{r} \right\rangle, 
\end{align} 
the gravitational potential at the center is $V_g(0)=-GM\, m_a/R$.  Then Eq. (\ref{gravpot}) implies that $1\ge w(z)>0$, with $w(0)=1$.

Phenomenological models of halos are of the form $\rho(r)=\rho(0) f(r/R_s)$, where $R_s$ is a radial scale, differing from the one defined in Eq. (\ref{harmonic}). The concentration is defined as $c=r_{\rm vir}/R_s$, the virial radius, $r_{\rm vir}$, is usually defined as the radius of a sphere containing the total mass of the halo and densities are cut off at $r_{\rm vir}$.

Using our method, all densities vanish at some finite value of $z=z_{\rm vir}$ which we identify with the scaled virial radius, $z_{\rm vir}=r_{\rm vir}/R$ . In phenomenological models, the scales of models, e.g. concentration, are not universal between models.  As Eq. (\ref{harmonic}) implies $ \langle 1/z\rangle=1$, we can give a universal definition to the concentration $c=z_{\rm vir}$.

Note that $R$ is of the same order of magnitude but larger than the scaling parameter of phenomenological models (\cite{Navarro:1995iw,Burkert:1995yz}) and unlike those, {\em model independent}. That can be shown by calculating the harmonic average, Eq. (\ref{harmonic}), in those models. The harmonic radius of a NFW halo is 
\begin{equation}\label{RNFW} 
R=R_{\rm NFW}\left[\left(1+\frac{1}{c_{\rm NFW}}\right)\log(1+c_{\rm NFW})-1\right]>R_{\rm NFW}, 
\end{equation} 
where $R_{\rm NFW}$ is the scaling radius of the NFW halo, and $c_{\rm NFW}$ is the NFW concentration. Similarly, the harmonic radius of the Burkert halo is 
\begin{equation}\label{RBurkert} 
R=R_{\rm B}\frac{\log(1+c_{\rm B}^2)+2\log(1+c_{\rm B})-2 \tan^{-1}(c_{\rm B})}{\log(1+c_{\rm B}^2)-2\log(1+c_{\rm B})+2 \tan^{-1}(c_{\rm B})}>R_{\rm B}, 
\end{equation} 
where $R_{\rm B}$ is the scaling radius of the Burkert halo, and $c_{\rm B}$ is the Burkert concentration.

Another interesting property of the gravitational scaling function, $w(z)$ is related to its behavior at the virial radius. Virial radius is defined by $w(z_{\rm vir})=0$. Taking Eq. (\ref{gravpot}) at $z=z_{\rm vir}$ and using Gauss's theorem we obtain 
\begin{equation}\label{virial}
V_g(z_{\rm vir})=-\frac{GM\, m_a}{r_{\rm vir}}=-\frac{GM\, m_a}{R\,z_{\rm vir}}=-\frac{GM\, m_a}{R}w(z_{\rm vir}), 
\end{equation} 
with the implication $w(z_{\rm vir})=1/z_{\rm vir}$.

We also define rescaled dimensionless wave functions, with unit normalization, as 
\begin{equation} 
\phi_{nl}(z)=R^{-3/2} N_{nl}^{1/2} \psi_{nl}(r). 
\end{equation} 
Writing Eq. (\ref{scheq}) in terms of rescaled wave functions we obtain 
\begin{equation}\label{Etoepsilon} 
\epsilon_{nl}\phi_{\epsilon l} =\frac{1}{S}\left(-\phi_{n l}''-\frac{2}{z}\phi_{ nl}+\frac{l(l+1)}{z^2}\phi_{nl}\right)-w(z)\phi_{\epsilon l}, 
\end{equation} 
where 
\begin{equation} \label{Etoepsilon2}
\epsilon_{nl}=\frac{1}{S}2m_a R^2 E_{nl} 
\end{equation} 
and where the dimensionless halo size parameter is 
\begin{equation}\label{Sdef} 
S=2MG R \,m_a^2. 
\end{equation} 
As the expectation value of the kinetic term is positive, Eq. (\ref{Etoepsilon}) implies that the range of the scaled energy parameter is $-1 \leq \epsilon\leq0$.

Rough estimates show that $S\gg 1$ even for moderate size galaxies, even if the ULA mass is as small as $m_a=10^{-22}$ eV. For the Milky Way $S_{MW}\gtrsim 10^5$. As a contrast to the galactic halo, $S=O(1)$ for a soliton, or axion star. 

Using Eq. (\ref{Etoepsilon}) we estimate the number of bound states. The depth of the rescaled potential well is $w(0)=1$ and as we see later the spectrum of $\epsilon_{nl}$ fills most of the interval $-1\leq \epsilon\leq 0$. As we will see in the next section, WKB quantization implies that $\epsilon$ is quantized as $\epsilon \sim - n/\sqrt{S}$, where $n$ is the principal quantum number. Then the principal quantum number takes up to $O(\sqrt{S})$ different values. Including states with nonzero angular momentum, we estimate that the number of energy levels in the potential well is $O(\sqrt{S}$, a very large number. We find no reason why most of those states would not be occupied by the astronomical number of ULAs. In a previous study \cite{Bernal:2009zy,PhysRevD.101.081302}, only a small number of excited states were considered.

\section{WKB approximation} 
Eq. (\ref{Etoepsilon}) lends itself to a perturbative WKB expansion in $\delta=1/\sqrt{S}$ \cite{book} which gives a solution to the differential equation, 
\begin{align} 
\phi_{nl}(z)={\text{Exp}}\left[\frac{1}{\delta}\sum_{k=0}^\infty\delta^k P_k\right], 
\end{align} 
where for the two independent solutions in the allowed, oscillating region are
\begin{align} 
P_0&=\pm\, i\,\int_{z_{\rm min}}^z dz'\sqrt{F_{nl}(z')},\\ P_1&=\log\left[\frac{{\cal{N}}_{nl}}{z F_{nl} (z)^{1/4}}\right] 
\end{align} 
while terms $P_2,...$ are of $O(S^{-1/2})$, and negligible. Eq.(\ref{Etoepsilon}) is linear, so any linear combination of solutions is admissible.The unique solution satisfying boundary conditions, finiteness at the turning points, which are zeros of $F_{nl}[z]$, is
\begin{align}\label{wave} 
\phi_{nl}=\frac{{\cal{N}}_{nl}}{z F_{nl} (z)^{1/4}}\sin\left(\sqrt{S}\int_{z_{\rm min}}^z dz'\sqrt{F_{nl}(z')}\right), 
\end{align} 
where  $F_{nl} (z)$ is 
\begin{equation}\label{Fdef} 
F_{nl} (z)= w(z)+\epsilon_{nl}-\frac{l(l+1)}{z^2\,S}. 
\end{equation} 
Note that wave functions $\phi_{nl}(z)$ are normalized to 1. Multiplier ${\cal N}_{nl}$ is introduced to ensure the correct normalization of wave functions, 
\begin{align}\label{norm} 
{\cal N}_{nl}^{-2}&=2\pi \int_{z_{\rm min}}^{z_{\rm max}}\frac{dz}{F_{nl} (z)^{1/2}}.
\end{align}

In the classically forbidden region the wave function decreases exponentially as $\phi\propto \exp(-\sqrt{S} c)$, where $c$ is finite as $S\to\infty$.. In the limit $S\to\infty$ the approximate solution Eq. (\ref{wave}) becomes exact. As usual, $z_{\rm min}$ and $z_{\rm max}$ are the turning points where $F_{nl}=0$ (\ref{wave}) is also known as the Wentzel ans\"{a}tz to the WKB solution of Eq. (\ref{Etoepsilon}).

Noting that in Eq. (\ref{wave}), factors other then the exponential function vary slowly as function of $n$ and $l$, the quantization condition for energy eigenvalues $\epsilon_{nl}$ can be read off from Eq. (\ref{wave}) \cite{PhysRev.51.669,PhysRevA.53.3798}: 
\begin{equation}\label{quant} 
\int_{z_{\rm min}}^{z_{\rm max}}\sqrt{F_{nl}}=\frac{n}{\sqrt{S}}\pi. 
\end{equation}

The principal quantum number, $n$, and the orbital quantum, $l$, only appear in combinations $\nu=n/\sqrt{S}$ and $\lambda=(l+1/2)/\sqrt{S}$.\footnote{We use Langer's method to define $\lambda$ \cite{PhysRev.51.669}\cite{PhysRevA.53.3798}.} As the separation of subsequent values of quantum numbers $\nu$ and $\lambda$ vanishes as $S$ increases, they will be replaced by continuous variables. That replacement will facilitate the construction of the density as a function of $w(z)$ in the next section. Using the continuous quantum numbers, we can write the wave function as 
\begin{align}\label{wave2} 
\phi_{\epsilon\lambda}=\frac{{\cal{N}}_{\epsilon\lambda}}{z F_{\epsilon\lambda} (z)^{1/4}}\sin\left(\sqrt{S}\int_{z_{\rm min}}^{z}dz'\sqrt{F_{\epsilon\lambda}(z')} \right), 
\end{align}

In Eq. (\ref{wave2}) we use an unique connection between quantum numbers $\nu$ and $\epsilon$ through quantization condition (\ref{quant}), as it will be explained int the next section.

\section{Construction of the halo} \label{sec:construct}
The rescaled wave function of the halo is obtained from Eq. (\ref{totalwave}) as 
\begin{equation}\label{totalwave2} 
\Phi(z,t)=\sum_{nlm} \phi_{nl}(z) Y_{lm}(\theta,\phi)e^{i(\alpha_{nlm}+\epsilon_{nl}t)}. 
\end{equation} 
Then the total density, averaged over rotations and random phases, and normalized to 1, is 
\begin {equation}\label{density2} 
\rho(z)=|\Phi(z,t)|^2\simeq\sum_{nl}(2l+1)\frac{M_{nl}}{M}|\phi_{nl}|^2. 
\end{equation}

Consider now that wave functions depend on quantum numbers $n$ and $l$ only through combinations $\nu=n/\sqrt{S}$, in Eq. (\ref{quant}) and $\lambda=\sqrt{l(l+1)}/\sqrt{S}$, in Eq. (\ref{Fdef}). The rescaled quantum numbers, $\nu$ and $\lambda$, have finite ranges and become dense in those ranges as $S\to\infty.$ Therefore, the error of turning summations over $n$ and $l$ into integrations over $\nu$ and $\lambda$ is only of $O(1/\sqrt{S})$ and negligible for almost all galaxies. Then we obtain 
\begin{align}\label{density3} 
\rho(z)=S^{3/2}\int d\nu\int d\lambda^2\frac{{\cal{N}}_{\epsilon\lambda}^2}{z^2 F_{\epsilon\lambda}^{1/2}}b(\epsilon,\lambda), 
\end{align} 
where dimensionless, continuous distribution function $b(\epsilon,\lambda)$ {\em interpolates distribution function} $M_{nl}/M$. It is normalized as 
\begin{align}\label{distribution} 
S^{3/2}\int d\nu\int d\lambda^2 b(\epsilon,\lambda)=1. 
\end{align}

The total variation of $\epsilon$ is 1, or $1+\epsilon_c$, in case there is a gap of size $|\epsilon_c|$. There are $O(\sqrt{S})$ energy levels, implying that the discrete values of $\epsilon$ are dense on interval $(-1,0)$. Wave functions depend explicitly on $\epsilon$ only, therefore we change integration variable $\nu$ to $\epsilon$. We use the quantization condition (Eq. (\ref{quant})), written in terms of continuous quantum numbers, 
\begin{equation}\label{quant1} 
\nu\,\pi=\int_{z_{\rm min}}^{z_{\rm max}}dz\sqrt{w[z]+\epsilon-\frac{\lambda^2}{z^2}}, 
\end{equation} 
to find the appropriate Jacobian. Taking the derivative of Eq. (\ref{quant1}) with respect to $\epsilon$ at constant $\lambda$, we note that terms coming from differentiating the integral with respect to boundary values $z_{\rm min}$ and $z_{\rm max}$ vanish. Then using Eq. (\ref{norm}) we obtain 
\begin{equation}\label{deriv2} 
\frac{d\nu}{d\epsilon}\pi=\frac{1}{2}\int_{z_{\rm min}}^{z_{\rm max}}\frac{dz}{\sqrt{w(z)+\epsilon-\frac{\lambda^2}{z^2}}}=\frac{1}{4 \pi}{\cal N}_{nl}^{-2}, 
\end{equation} 
Substituting Eq. (\ref{deriv2}) into Eq. (\ref{density3}) we arrive at our final result for the density: 
\begin{align}\label{density4} 
\rho(z)=\frac{S^{3/2}}{4\pi^2}\int_{-w(z)}^{\epsilon_{\rm max}} d\epsilon\int_0^{z^2(w(z)+\epsilon)} \frac{d\lambda^2 b(\epsilon,\lambda)}{z^2 \sqrt{w[z]+\epsilon-\frac{\lambda^2}{z^2}}}, 
\end{align} 
where $\epsilon_{\rm max}\le0$ is an admissible integration constant allowing for the existence of an energy gap. If the distribution function $b(\epsilon,\lambda)$ is independent of the angular momentum quantum number, $\lambda$, integration over $\lambda$ yields 
\begin{align}\label{density5} 
\rho(z)&=\frac{S^{3/2}}{2\pi^2}\int_{-w(z)}^{\epsilon_{\rm max}} d\epsilon \sqrt{w[z]+\epsilon}\,b(\epsilon)=\rho(0)\frac{\int_{-w(z)}^{\epsilon_{\rm max}} d \epsilon \sqrt{w(z)+\epsilon}\,b(\epsilon)}{\int_{-1}^{\epsilon_{\rm max}}d\epsilon \sqrt{1+\epsilon}\,b(\epsilon)}. 
\end{align}

Even if $b(\epsilon,\lambda)$ depends on $\lambda$ it is expected that that an expansion with respect to $\lambda^2$ converges rapidly. Then, using the expansion 
\begin{align} 
b(\epsilon,\lambda)=\sum_{k=0}b(\epsilon)^{(k)}\lambda^{2k} 
\end{align} 
we can integrate the series with respect to $\lambda^2$, term by term to obtain 
\begin{align} \label{density52} 
\rho(z)=\frac{\rho(0)}{2}\frac{\sum_k z^{2k}\frac{\sqrt{\pi}\Gamma(1+k)}{\Gamma(3/2+k)}\int_{-w(z)}^{\epsilon_{\rm max}}d\epsilon \, b^{(k)}(\epsilon)(w(z)+\epsilon)^{k+1/2}}{\int_{-1}^{\epsilon_{\rm max}}d\epsilon\sqrt{1+\epsilon}\,b(\epsilon)} 
\end{align} 
Using Eqs. (\ref{density5}) or (\ref{density52}), the Poisson equation,
\begin{equation}\label{poissonw} 
w''(z)+\frac{2}{z}w'(z)=-4\pi \rho(z) 
\end{equation} 
becomes a differential equation for $w(z)$. Since $w(0)=1$ and $w'(0)=0$, to insure that the density is regular at the center (the exception is the case when density is replaced by the NFW profile), after providing a distribution function there are no undetermined integration constants, other than $\epsilon_{\rm max}$ and $\rho(0)$. As we will see in Sec. \ref{sec:examples}, for some distribution functions $b(\epsilon,\lambda)$ the density (Eq. (\ref{density4})) or (\ref{density5}), as appropriate, can be analytically integrated.

The strategy for solving Eq. (\ref{poissonw}) is simpler for Eq. (\ref{density5}) then for Eq. (\ref{density52}), because after rescaling $z$ as $x=z\,\sqrt{\rho(0)}$, eliminating the central density from the equation, we find a unique solution at fixed $\epsilon_{\rm max}$. Among others, we find $\langle 1/x\rangle$. Then we can restore the original scaling variable $z= x \langle 1/x \rangle$, which is true because $\langle 1/z\rangle=1$. We also find $\rho(0)=\langle 1/x\rangle^2.$

It is more complicated to find the solution of Eq. (\ref{poissonw}) in the case when the energy spectrum is not degenerate. Then, due to the explicit dependence of $\rho(z)$ on $z$, coordinate $z$ cannot be eliminated from Eq. (\ref{density52}) by a simple rescaling. In that case, rather than rescaling $z$ we need to search for an appropriate value of $\rho(0)$ at fixed $\epsilon_{\rm max}$, such that $\rho(z)$ is normalized to 1.

Dynamical simulations generate numerical distribution functions, as well \cite{Lin:2018whl}. A time dependent version of this work, to be published, can potentially do that as well. One advantage of our analytic approximation method is scalability: dynamical simulations are limited by computational power to relatively small galactic halos.

Finally, we note that rotation curves have simple scale invariant representations in terms of our scaling variable and scaling function. Up to an overall factor of dimension of velocity, $v_0$, 
\begin{equation} 
v(r)=v_0\sqrt{-z\,w'(z)}. 
\end{equation}

\section{Virial theorem in WKB} 
Simulations show that, in agreement with expectations, the halo is virialized after collapse \cite{Schive:2014hza}\cite{Schwabe:2016rze}. Using {\em stationary} WKB wave functions the halo does not automatically satisfy the virial theorem. However, a time dependent generalization of the WKB approximation is expected to to converge to a virialized state. Such a generalization is deferred to a future publication. 

The virial theorem for a self-gravitating system, without contact interactions, is $2 K +E_g=0$, where $K$ is the total kinetic energy and $V_g$ is the total gravitational energy. Using the definitions of the previous sections, they are 
\begin{align}\label{KV} 
K&=\frac{1}{2m_a^2}\sum_{nl}(2l+1)\int d^3r\left[|\psi_{nl}'(r)|^2 +\frac{l(l+1)}{r^2}|\psi_{nl}(r)|^2 \right],\\ E_g&=\frac{1}{2m_a}\sum_{nl}(2l+1)\int d^3r|\psi_{nl}(r)|^2 V_g(r), 
\end{align} 
where $V_g(r)$ has been defined in Eq. (\ref{Vg}). Using the dimensionless wave functions and rescaled radial parameter, $z=r/R$, we can rewrite $K$ and $V$ as 
\begin{align}\label{KV2} 
K&=\frac{1}{2m_a^2R^2}\sum_{nl}(2l+1)M_{nl}\times\int d^3z\left[\phi_{nl}'[z]^2+\frac{l(l+1)}{z^2}|\phi_{nl}(z)|^2\right],\\ E_g&=-\frac{MG}{2R}\sum_{nl}(2l+1)M_{nl}\int d^3z|\phi_{nl}(z)|^2 w(z) 
\end{align}

Finally, using Eq. (\ref{wave2}), introducing continuous energy and angular momentum variables as in Eq.  (\ref{density3}), and expressing $G$ by size parameter $S$, we obtain 
\begin{align} 
K&=2C\int dz \int_{-w(z)}^{\epsilon_c} d\epsilon\int_0^{z^2[w(z)+\epsilon]}d\lambda^2 b(\epsilon,\lambda)\left[\sqrt{w(z)+\epsilon-\frac{\lambda^2}{z^2}}+\frac{\lambda^2}{z^2\sqrt{w(z)+\epsilon-\frac{\lambda^2}{z^2}}}\right],\\ E_g&=-C\int dz\int_{-w(z)}^{\epsilon_c} d\epsilon\int_0^{z^2[w(z)+\epsilon]}d\lambda^2 b(\epsilon,\lambda)\frac{w(z)}{\sqrt{w(z)+\epsilon-\frac{\lambda^2}{z^2}}}, 
\end{align} 
where $C=S M/(4m_a^2R^2)$, is a constant and $\epsilon_c\leq0$ is an admissible energy cutoff parameter. Then the virial theorem applied to our system becomes independent of all dimensional parameters and of dimensionless size parameter, $S$. It reads as 
\begin{align}\label{virial1} 
\int_9^{z_{\rm vir}} dz\, z^2\int_{-w(z)}^{\epsilon_c} d\epsilon\int_0^{z^2[w(z)+\epsilon]}d\lambda^2 \frac{b(\epsilon,\lambda)(3w(z)+4\epsilon)}{\sqrt{w(z)+\epsilon-\frac{\lambda^2}{z^2}} }=0. 
\end{align} 
In the particular case when distribution function, $b(\epsilon\lambda)\to d_\epsilon$ is independent of $\lambda$, we can integrate over $\lambda$ and the virial theorem becomes 
\begin{align}\label{virial2} \int_9^{z_{\rm vir}} dz\, z^2\int_{-w(z)}^{\epsilon_c} d\epsilon\,b(\epsilon)\sqrt{w[z]+\epsilon}(3w(z)+4\epsilon)=0. 
\end{align}

Suppose now that an ansatz for the distribution function $b(\epsilon,\lambda)$ depends on a free parameter. A simple example is, $\epsilon_{\rm max}$, defining a gap in the energy spectrum, $b(\epsilon,\lambda)=0$ for $\epsilon_{\rm max}<\epsilon<0$. Another possibility is that a background density, $\rho_0$, of undetermined size is subtracted from the density. Then, using the procedure described in Sec. \ref{sec:construct}, we find the numerical gravitational scaling function, $w(z)$, as a function of $\epsilon_{\rm max}$ or $\rho_0$. Substituting into Eq. (\ref{virial1}) allows us to fix $\epsilon_{\rm max}$ or, alternatively, $\rho_0$, which we will demonstrate in the next section. If there is no choice for the free parameter to satisfy the condition given by Eq. {\ref{virial1}}, then the distribution function $b(\epsilon,\lambda)$ does not allow for the system to reach dynamical equilibrium and is not physically acceptable.

\section{Examples for halos} \label{sec:examples}
To complete the calculation of the density (Eq. (\ref{density4})) we have to know the energy distribution function $b(\epsilon,\lambda)$. In this section we will explore a variety of physical choices for $b(\epsilon,\lambda)$ to check whether they provide acceptable density distributions. In particular, we will pay particular attention to whether the obtained density distribution are in agreement with the following general features of observational data and simulations.

\begin{itemize} 
\item It is generally accepted \cite{Schive:2014dra}\cite{Hui:2016ltb}\cite{Schwabe:2016rze} that after its collapse, at least asymptotically in time, the halo is virialized, i.e. satisfies virial condition, $2K+E_g=0$. In simulations the system is driven towards dynamical equilibrium. As we pointed out earlier, unlike in simulations, $b(\epsilon,\lambda)$ must have a parameter, which can be chosen such that the condition is satisfied. 
\item Another important feature is that galaxies and galaxy clusters are coming in a very wide range of sizes. As condensation $c$ controls the size of the halo, general solutions must have solutions for a large range of $c$. 
However, not all density distributions satisfy this criterion, as an example, a model which we examined in a previous work \cite{Street:2021qxl}.
\item Popular models for galactic halos predict that the rotation curves are independent of the concentration. Though all observed rotation curves are close to being flat, significant variations have been observed \cite{Zintner:2022zds}. This is especially important if one tries to describe a wide variety halos, including galaxy clusters. 
\end{itemize}

\subsection{Thermal equilibrium} 
Recently, we applied the WKB expansion method to investigate a system in thermal equilibrium \cite{Street:2021qxl}. We give a simplified account of those investigations, below. Neglecting gravitational interactions, the system is treated as an ideal Bose-Einstein gas. The number of particles on energy levels $E_n$ is $N_n$. Then the partition function for the for the system is 
\begin{equation}\label{partition} 
e^{-\beta\, g}=\prod_n\frac{1}{1-e^{-\beta(E_n-\mu)}}. 
\end{equation} 
The average value of $N_n$ is 
\begin{equation} 
\langle N_n\rangle=\frac{1}{e^{\beta(E_n-\mu)}-1} 
\end{equation} 
The occupation number, $\langle N_n\rangle$, is enormous in every state, $n$.  As a result, every exponent must satisfy the inequality $1 \gg \beta\left( E_n - \mu \right) >0$.  Consequently, we can expand them, keeping terms to linear order, to arrive at the Rayleigh-Jeans limit of Bose-Einstein statistics.
\begin{equation}\label{RayleighJeans} 
\langle N_n\rangle=\frac{1}{\beta(E_n-\mu)} 
\end{equation} 
Using Eq. (\ref{Etoepsilon}) and rescaling $\beta$ and $\mu$ as 
\begin{align} 
\beta&=\frac{2m_aR^2}{S}\tilde{\beta},\\ \nonumber & \mu=\frac{S}{2m_aR^2}\tilde{\mu} 
\end{align} 
we substitute into Eq. (\ref{density5}), to get 
\begin{align}\label{eqden1} 
\rho(z)=\rho_0\left(\int_{-w(z)}^0\frac{\sqrt{w(z)+\epsilon}}{\epsilon-\tilde{\mu}}d\epsilon-\rho_B\right), 
\end{align} where 
\begin{equation*} 
\rho_0=\frac{S^{3/2}}{4\pi^2\tilde{\beta}} 
\end{equation*} 
and $\rho_0\rho_B$ is a background density term. $\rho_B$ is considered to be an adjustable parameter.

Now remember that the ground state corresponds to $\epsilon=-1$, Then the physical range of the chemical potential is $\tilde\mu\leq-1$. During the collapse of the halo the system cools and the chemical potential increase towards its critical value, $\tilde\mu=-1$, where Bose-Einstein condensation starts. Integrating density $\rho(z)$ and setting $\tilde\mu=-1$ we obtain Poisson equation 
\begin{align}\label{thermalPoisson1} 
&w''(z)+\frac{2}{z}w'(z)=-8\rho_0\pi\\ &\nonumber\left[\tan^{-1}\left(\sqrt{\frac{w(z))}{1-w(z)}}\right)\sqrt{1-w(z)}-\sqrt{w(z)}-\frac{\rho_B}{2}\right]. 
\end{align} 
To facilitate the numerical solution of Eq. (\ref{thermalPoisson1}) we rescale coordinate $z\to x/\sqrt{8\rho_0\pi}$. Then integrating the Poisson equation at a series of choices for background density parameter $\rho_B$, we find that at $\rho_B=0.017$, the virial condition (Eq. (\ref{virial2})) is satisfied. As at that choice of $\rho_B$ $w(x_{\rm vir})=0$, where $x_{\rm vir}\simeq4.74$. We also calculate the concentration, 
\begin{equation*} 
c=\left\langle \frac{1}{x}\right\rangle x_{\rm vir}\simeq2.78 
\end{equation*} 
A similar value is obtained for $c$ if we modify the distribution function to take into account gravitational interactions. That can be done by replacing energy level $E_n$ by $H_n$, the contribution of that energy level to the conserved Hamiltonian. Since observations imply $c\gtrsim 10$, the final state of the collapse cannot be in dynamical equilibrium and thermal equilibrium at the same time.

\subsection{The King model} 
The King model \cite{1966AJ.....71...64K} is based on the classical kinetic theory of collision-free self-gravitating particles, assuming Maxwell velocity distribution cut off at an escape velocity. Then the equilibrium energy distribution takes the form 
\begin{equation}\label{King} f_{\rm King}=
\begin{cases}
A \left(e^{-\beta(E-E_c)}-1\right)& \text{if } E\leq E_c\\ 0&\text{otherwise.} 
\end{cases} 
\end{equation} 
In a recent paper, \cite{Lin:2018whl}, galactic halos were constructed, using the King distribution \cite{1966AJ.....71...64K} (and other classical distribution functions, like the fermionic King model and the Ossipov-Merritt model \cite{1979PAZh....5...77O,1985AJ.....90.1027M}), combined by wave functions of excited states of Eq. \ref{scheq} obtained by simulations. The halos were compared to profiles obtained by simulations of collapsing systems of ultralight bosons, with excellent agreement. Unlike in \cite{Lin:2018whl}, we do not add a contribution for the condensate at the core of the density distribution, because our aim is to compare the densities predicted by the King model to phenomenological models, rather than the results of simulations.  However, our result are in excellent agreement with simulations outside of the $z\lesssim 0.1$ region where the soliton core dominates the density distribution.

We will use our analytic WKB wave functions (Eq. (\ref{wave2})) combined with the King distribution function to construct halos, which reduces the number of parameters used in the construction.  The fermionic King model, also used in \cite{Lin:2018whl} give slightly better fits, as it has an extra adjustable parameter. As we will see below, using WKB wave functions, rather than simulations, has the advantage of scalability, which allows us to find new features when investigating larger systems.

Adapting Eq. (\ref{King}) to WKB wave functions we write Eq. (\ref{density5}) as 
\begin{align}\label{density6} 
\rho(z)=\frac{A\,S^{3/2}}{4\pi^2}\int_{-w(z)}^{\epsilon_c}d\epsilon\left(e^{-\beta(\epsilon-\epsilon_c)}-1\right)\sqrt{w[z]+\epsilon}, 
\end{align} 
Scaling out coefficient $A\,(S/\beta)^{3/2}/4\pi^2$ from the density and rescaling $z\to x$, as explained in Sec. \ref{sec:construct}, we integrate over $\epsilon$ to get the Poisson equation 
\begin{align}\label{qequation} 
q''(x)&+\frac{2}{x}q'(x)\\ \nonumber&=-3\left[e^{q(x)}\text{erf}\left(\sqrt{q(x)}\right)-6\sqrt{q(x)}-4q(x)^{3/2}\right], 
\end{align} 
where 
\begin{equation}\label{qbeta} 
q(x)=\beta(w(z)+\epsilon_c), 
\end{equation} 
and where 
\begin{equation} 
x=z \frac{A^{1/2}S^{3/2}}{2\sqrt{6}\pi\beta^{1/4}}. 
\end{equation}

\begin{figure}[hbt!] 
\includegraphics[width=0.5\textwidth]{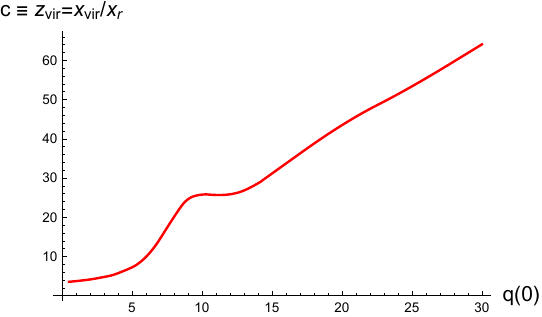} 
\caption{\label{plotc} Virial radius as a function of $q(0)$} 
\end{figure}

The only adjustable parameter in the Poisson equation (Eq. (\ref{qequation})) is $q(0)=\beta(1+\epsilon_c)$, which we took be to a number of values.  We found that the concentration, $c  = z_\text{vir} = x_\text{vir}\langle 1/x\rangle$, plotted in Fig. \ref{plotc}, is a rapidly rising function of $q(0)$.  The implication is that the King model can describe halos of markedly different sizes.

\begin{figure}[hbt!] 
\includegraphics[width=0.8\textwidth]{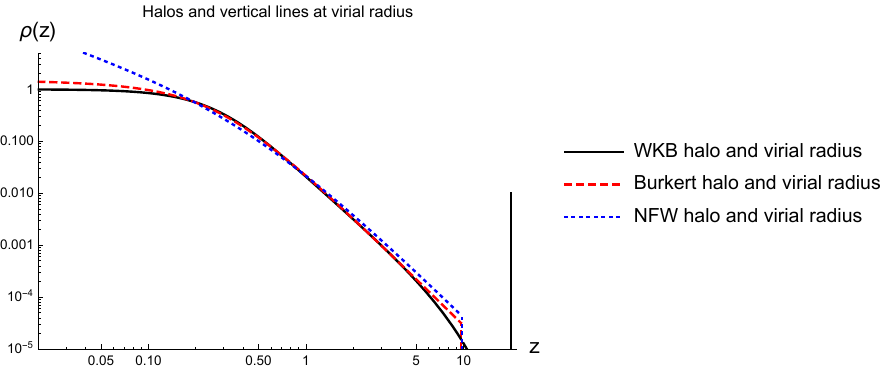} 
\caption{\label{eight}King model density distribution compared to NFW and Burkert profiles at $q(0)=8.$} 
\end{figure}

We found that in the range $7\lesssim q(0)\lesssim 10$, the King model density distributions are quite close to NFW or Burkert distributions well in the region where the bulk of the density is distributed, $0.1\lesssim z\lesssim 10$. As an example, we plot the WKB density distribution, $\rho(z)$, corresponding to initial condition $q(0) =8$ in Fig. \ref{eight}.  We switch to the physical, rescaled radial coordinate, $z$ and choose central density $\rho(0)=1$ in the plot.  NFW and Burkert profiles, the core density and core radius parameters of which were fitted to the numerically calculated halo, are also plotted.  We found fitted parameters $R_{\text{NFW}}= 0.437R$, and $R_\text{Burkert} = 0.269R$, where $R$ is the harmonic radius of the WKB solution.  Then, we were able to calculate the virial radii and concentrations of those profiles using Eqs. (\ref{RNFW}) and (\ref{RBurkert}).  We found $c_{\text{NFW}} = 22.3$ and $c_{\text{Burkert}} = 35.7$ and physical virial radii,
\begin{align}
z_\text{NFW} &= c_\text{NFW} R_\text{NFW} / R = 9.78,
\nonumber \\
z_\text{Burkert} &= c_\text{Burkert} R_\text{Burkert} / R = 9.74,
\end{align}
which are quite close to each other.  The virial radius of the WKB halo is $z_\text{vir} = 20.1$, as it does not have a sharp cutoff.  However, in the region $10 < z < 20$, $\rho(z) = \mathcal{O}\left( 10^{-5} \right)$.

\begin{figure}[hbt!] 
\includegraphics[width=0.7\textwidth]{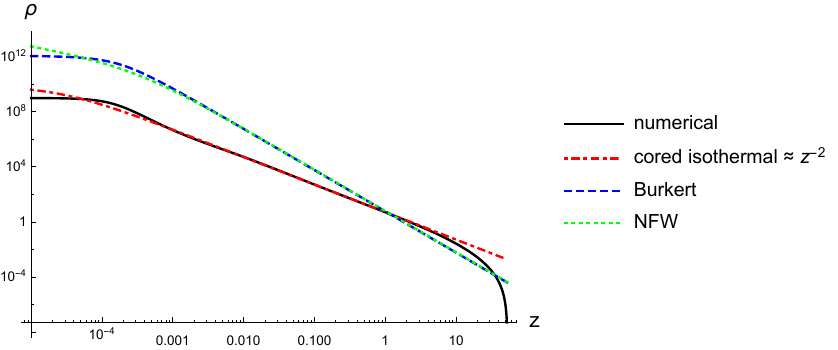} 
\caption{\label{25}King model density distribution compared to isothermal, NFW and Burkert profiles at $q(0)=25.$} 
\end{figure}

Next, we examine the predictions of the King model at a larger virial radius, in the range where $c$ is a linearly rising function of $q(0)$, as it can be seen in Fig. \ref{plotc}. Very likely that range corresponds to galaxies larger than the Milky Way, or galaxy clusters, which are currently not accessible for simulations.  At $q(0)\gtrsim12$ density profiles are very different from those at $q(0)=8$, plotted in Fig. \ref{25}. In fact, they are in excellent agreement with a slightly cored isothermal profile, with behavior $\rho(z)\propto z^{-2}.$ We plot $\rho(z)$ at $q(0)=25$ as a function of $z$ in Fig. \ref{25}, along with a weakly cored pseudo-isothermal profile, an NFW profile, and a Burkert profile. All profiles are scaled to match at $z=1$, which corresponds to the harmonic average radius of the system. Note that the bulk of contributions to the total mass come from the range $0.01 \lesssim z\lesssim 10$.

Clearly, the King model profile is in excellent agreement with the isothermal profile  \cite{May:2022may}\cite{Jim:2003jvt}\cite{Shu:2008cis}\cite{Bro:2009.bs}, while the NFW and Burkert profiles are not.

Lin et. al. \cite{Lin:2018whl} generated two halos labeled A and B, for which they fit energy cutoff parameter, $E_c$ and scale parameter, $\beta_{\rm sim}$, defining the King model. We wish to compare those with parameters we use to construct WKB halos. However, the corresponding values of $\beta_{\rm WKB}$ and $\epsilon_c$ cannot be directly compared with the simulations of \cite{Lin:2018whl}, because we rescaled energy $E\to\epsilon$ and, consequently, our $\beta$ value is also rescaled, though rescaling leaves the dimensionless quantity $E_c\beta_{\rm sim}=\epsilon_c\beta_{\rm WKB}$, unchanged. To check that equality, we need to use the relationship (Eq. (\ref{qbeta})), which implies of $q(0)=\beta_{\rm WKB}(1+\epsilon_c)$. Note now, that cutoff parameter $\epsilon_c$ has not been fixed in previous calculations. In other words, $\epsilon_c$ is a free parameter. 

Consider now that the halos generated in \cite{Lin:2018whl} are virialized, while there is no reason why our halos would in general be. In fact, calculating (\ref{virial2}) as 
\begin{align} 
2\,K +E&\propto \int_0^{z_{\rm vir}} dz\,z^2\int_{-w(z)}^{\epsilon_c}d\epsilon\sqrt{w(z)+\epsilon}\\ \nonumber&\times\left(e^{-\beta(\epsilon-\epsilon_c)}-1\right)\left(\epsilon+\frac{3}{4}w(z)\right) 
\end{align} 
we find that at every $q(0)$ we have $\langle 2\,K +E\rangle=a+b\, \beta\,\epsilon_c$ with varying fixed values for $a$ and $b$. Then the vanishing of $\langle2\,K +E\rangle$ fixes the combination $\beta\epsilon_c$ at every choice of $q(0)$. 

\begin{figure}[hbt!] 
\includegraphics[width=0.7\textwidth]{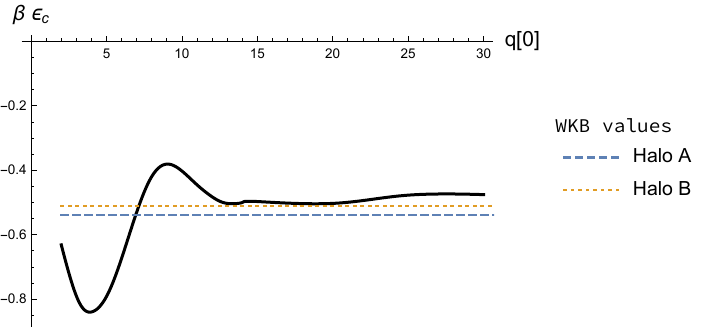} 
\caption{\label{betaplot}}Plot of King model values of $\beta\,\epsilon_c$ along with those of halos A and B obtained in simulations of \cite{Lin:2018whl}. 
\end{figure}

We plot the combination $ \epsilon_c\beta$ as function of $q(0)$ along with the values of $E_c\beta_{\rm sim}$ of halos $A$ and $B$, $-0.539$ and $-0.51$, of \cite{Lin:2018whl}, in Fig. \ref{betaplot}. Those combinations are equal after rescaling $E_c$ and $\beta$. The deviation of values of $\beta\,\epsilon_c$ between those found in our WKB calculation and those in simulations, are only a few percent at the relevant values of $q(0)$.

\section{Summary}
We have investigated the possibility that galactic dark matter haloes could be made up of ultra-light bosons with wave functions that satisfy nonlinear Schrodinger-Poisson equations. We have found eigenstates and eigenvalues of the Schrodinger equation using the  WKB approximation. The approximation method becomes more accurate as the galaxy's mass rises, however it has been demonstrated that even for smaller galaxies, the leading WKB approximation may yield accurate answers. The WKB approximation expresses the wave functions in terms of the gravitational potential. The energy levels were determined by the Bohr–Sommerfeld quantization condition, implied by the WKB method. To determine how we may better explain the data we have compared two an\"{a}tze for the level occupation number distributions, the Bose-Einstein distribution and the King model, at appropriate values of temperature and chemical potential. The modulus square of each wave function multiplied by the occupation number, summed over all the bound states, yielded the total density of particles. The mass density is therefore a function of the gravitational potential $V_g$ since the Poisson equation connects $V_g$ to the particle density. This procedure allowed us to obtain a differential equation for the gravitational potential.  Solving this equation enabled us to obtain the mass distribution of the ULDM model of a galaxy.  This technique can  easily be scaled up to model the DM halos of galaxies or galaxy clusters, well beyond the mass range covered by simulations.

When we used the Bose Einstein density distribution in our computation of the density profile we observed that the concentration parameter, $c$, of the resulting galaxy was smaller in magnitude than what is observed for most galaxies. When the King model particle distribution was utilized the concentration parameters, on the other hand, were within the permissible range for Milky way-like galaxies.

In future work we plan to study the dynamical collapse of boson clouds using  variational technique and study the times scales for collapse and decay in more detail. We will also investigate how the inclusion of self-interactions affects the the dynamics of halo. 
\newpage

 \begin{acknowledgments}
The authors are indebted to Joshua Eby for fruitful discussions. L.S. also thanks the Department of Physics at the
University of Cincinnati for financial support in the form of the Violet M. Diller Fellowship.
During part of this research L.S. was supported by the U.S. Department of Energy (DOE), Office of Science, Office of Workforce Development for Teachers and Scientists, Office of Science Graduate Student Research (SCGSR) program. The SCGSR program is administered by the Oak Ridge Institute for Science and Education (ORISE) for the DOE. ORISE is managed by ORAU under contract number DE-SC0014664. All opinions expressed in this paper are the
authors? and do not necessarily reflect the policies and views of DOE, ORAU, or ORISE.
Research of L.C.R.W. is partially supported by the US. Department of Energy grant DE-SC1019775.
\end{acknowledgments}
\bibliography{core_halo}
\bibliographystyle{JHEP}

\end{document}